\documentclass[lettersize,journal]{IEEEtran}
\usepackage{graphicx}
\usepackage{amsfonts}
\usepackage{subfigure}
\usepackage{float}
\usepackage{datetime2}
\usepackage{amsmath}
\usepackage{braket}
\usepackage{amssymb,cite}
\usepackage{enumitem}
\author{Kai Zheng, Miaowen Wen, \textit{Senior Member, IEEE}, Tianqi Mao, \textit{Member, IEEE},  Lixia Xiao, \textit{Member, IEEE}, 
	
	and  Zhaocheng Wang, \textit{Fellow, IEEE}
\thanks{
	{Kai Zheng and Miaowen Wen are with the School of Electronic and Information Engineering, South China University of Technology, Guangzhou 510640, China (Email:  eekaizheng@163.com, eemwwen@scut.edu.cn).}
	
	{Tianqi Mao is with the MIIT Key Laboratory of Complex-Field Intelligent Sensing, Beijing Institute of Technology,
		Beijing 100081, China, and also with theYangtze Delta Region Academy, Beijing Institute of Technology (Jiaxing), Jiaxing 314019, China (Email: maotq@bit.edu.cn).}
		
    {Lixia Xiao is with the Wuhan National Laboratory for Optoelectronics and
    the Research Center of 6G Mobile Communications, Huazhong University
    of Science and Technology, Wuhan 430074, China (Email: lixiaxiao@
    hust.edu.cn).}
    
    {Zhaocheng Wang is with the Department of Electronic Engineering, Tsinghua University, Beijing 100084, China, and also with the Shenzhen International Graduate School, Tsinghua University, Shenzhen 518055, China (Email:
    zcwang@tsinghua.edu.cn).}
}
}
\title{Channel Estimation for AFDM 
	
	With Superimposed Pilots}
\begin{document}
	\maketitle
	\begin{abstract}
        The recent proposed affine frequency division multiplexing (AFDM) employing a multi-chirp waveform has shown its reliability and robustness in doubly selective fading channels. In the existing embedded pilot-aided channel estimation methods, the presence of guard symbols in the discrete affine Fourier transform (DAFT) domain causes inevitable degradation of the  spectral efficiency (SE). To improve the SE, we propose a novel AFDM channel estimation scheme by introducing the superimposed pilots in the DAFT domain. An effective pilot placement method  that minimizes the channel estimation error is also developed with a rigorous proof. To mitigate the pilot-data interference, we further propose an iterative channel estimator and signal detector. Simulation results demonstrate that both channel estimation and data detection performances can be improved by the proposed scheme as the number of superimposed pilots increases.
	\end{abstract}
	
	\section{Introduction}
    Orthogonal frequency division multiplexing (OFDM) has been widely adopted in existing wireless communication systems. OFDM boasts excellent resistance to frequency-selective fading channels and the capability to achieve full diversity. However, in the next generation of wireless communication systems, the spectrum will expand into higher frequencies, and scenarios with high mobility (such as high-speed railway communication, V2X systems, and satellite communications) will become prevalent, resulting in  excessive Doppler shift effects. These inevitably induce strong time variance/time selectivity of the propagation channel, which disrupts the orthogonality of the OFDM subcarriers, leading severe inter-carrier interference. Furthermore, for the high-mobility scenario with multi-path effects, the communication signals have to suffer from both frequency and time selective channel fading, referred to as doubly selective fading.  To address this issue, one possible solution is to identify a new set of orthogonal bases for multi-carrier data transmission \cite{8869705}. 
    
    Recent researches have introduced several new modulation waveforms employing the novel orthogonal bases to combat doubly selective channel fading.
    Orthogonal time frequency space (OTFS) modulation \cite{7925924}, a novel two-dimensional modulation employing symplectic finite Fourier transform, outperforms OFDM in time-varying channels. In \cite{9392379},  the author rigorously derived that modulation in the delay-Doppler domain can be realized with a set of orthogonal bases using the ZAK representation. Additionally, affine frequency division multiplexing (AFDM) was proposed which is one-dimensional modulation in discrete affine Fourier transform (DAFT) domain \cite{10087310}. AFDM multiplexes a set of orthogonal chirp signals generated using DAFT with two carefully chosen parameters to outperform OFDM in time-varying channels. Both AFDM and OTFS are considered to be promising for implementation in  the next-generation communication systems.
	
	In wireless communication systems, accurate and timely channel estimation is crucial for data detection especially for doubly selective fading channels. There have been preliminary researches on channel estimation of the aforementioned novel modulation waveforms.  In~\cite{8671740}, the author proposed embedded pilot-aided channel estimation for OTFS, which utilized the guard symbols to mitigate mutual interference between pilot and data signals. Similarly, in \cite{10087310}, embedded pilot-aided channel estimation for AFDM was introduced, requiring less overhead of guard symbols compared to that for OTFS. However, in scenarios with large Doppler shift ranges or under low-latency requirement, the presence of guard symbols inevitably decreases the spectral efficiency (SE). To address this issue, the superimposed pilot scheme for OTFS was proposed in~\cite{9456894,10333832,9539066}. In \cite{9456894}, a single pilot symbol was superimposed in the OTFS frame, increasing the SE at the cost of reliability. Furthermore, the multiple superimposed pilot scheme was proposed for OTFS in  \cite{10333832} and \cite{9539066}, which outperformed the single pilot scheme.
	In the context of AFDM, the multiple embedded pilot-aided channel estimation was proposed to provide accurate channel state information at the receiver under multi-user scenarios~\cite{9880774}. However, this channel estimation method significantly degrades the SE due to usage of multiple guard symbols especially under massive-user communication scenarios~\cite{9880774}. 
	
	 To address the SE challenges in AFDM and different principles between OTFS and AFDM, we present a novel AFDM channel estimation approach by introducing the superimposed pilots. Specifically, the superimposed pilot philosophy is employed in the DAFT domain. Then the pilot placement strategy is optimized to achieve minimal mean estimation error with rigorous derivations. With such arrangement, we propose a novel AFDM channel estimation approach capable of capturing the Doppler shift, delay, and channel coefficient parameters of the time varying channel.

	\emph{Notations}: Vectors and matrices are represented by lowercase bold letters and uppercase bold letters, respectively. The superscripts $(\cdot)^*, (\cdot)^T$, and $(\cdot)^H$ denote the conjugate, transpose, and Hermitian operations, respectively. The $N\times N$ identity matrix is symbolized by $\mathbf{I}_N$. $\mathcal{CN}(\mu, \sigma^2)$ represents the complex Gaussian distribution with mean $\mu$ and variance $\sigma^2$. The operator $\left<\cdot\right>_N$ denotes the modulo operation with divisor~$N$. For any matrix $\mathbf{A}$, $\mathbf{A}(m,n)$ denotes the element in the $m$-th row and $n$-th column of matrix $\mathbf{A}$. $\text{Tr}\{\mathbf{A}\}$ represents the trace of the matrix $\mathbf{A}$. $\lfloor x\rfloor$ denotes the greatest integer number smaller than or equal to $x$. $\mathbb{E}\{\cdot\}$ denotes the expectation operator. The discrete Dirac function $\delta(x)$ equals~$1$ when $x=0$ and is zero otherwise. The operator diag($\mathbf{v}$) tramsfroms the vector $\mathbf{v}$ into a diagonal matrix.

	\section{Basic Concepts of AFDM}
	
	 Let $\mathbf{x}=[x_0,x_1,\ldots,x_{N-1}]^T$ denote the vector of $N$ phase shift keying (PSK) symbols in the DAFT domain. After mapping $\mathbf{x}$ to the time domain using the inverse DAFT, the modulated signal in the time domain is expressed as
	\begin{equation}
		\mathbf{s}=\boldsymbol{\Lambda}_{c_1}^H\mathbf{F}^H\boldsymbol{\Lambda}_{c_2}^H\mathbf{x},
	\end{equation}
	where ${\mathbf{F}}$ is the $N$-point discrete Fourier transform (DFT) matrix with ${\mathbf{F}}(m,n)=e^{-j2\pi mn/N}$,  $m, n\in\{1, \ldots, N\}$, and  $\boldsymbol{\Lambda}_{c}$ is given by
	\begin{equation}
		\boldsymbol{\Lambda}_{c}=\text{diag}\left(\left[1, e^{-j2\pi c}, \ldots, e^{-j2\pi c(N-1)^2}\right]^T\right).
	\end{equation}
	After the transmission of the time-domain signal over a linear time-varying channel
	\begin{equation} \label{channel}
		g_n(l)=\sum_{i=1}^Ph_ie^{-j2\pi f_in}\delta(l-l_i),
	\end{equation} 
	the time-domain signal $\mathbf{r}$ arrives at the receiver, whose $n$-th sample is given by
	\begin{equation}
		r_n=\sum_{l=0}^{+\infty} s_{n-l}g_n(l)+w_n,
	\end{equation}
	where $w_n\sim \mathcal{CN}\left(0, N_0\right)$  is additive white Gaussian noise  (AWGN). In (\ref{channel}), $P > 1$ is the number of channel paths, and $h_i$, $l_i$, and $f_i$ denote the channel coefficient, delay index, and Doppler shift associated with the $i$-th path, respectively.
	 
	 After the DAFT  operation, the time-domain received signal is transformed back to the DAFT domain as
	 \begin{align} \label{H_eff}
	 	\mathbf{y}=&\boldsymbol{\Lambda}_{c_2}\mathbf{F}\boldsymbol{\Lambda}_{c_1}\mathbf{r}\nonumber\\
	 	=&\sum_{i=1}^Ph_i\boldsymbol{\Lambda}_{c_2}\mathbf{F}\boldsymbol{\Lambda}_{c_1}\boldsymbol{\Gamma}_{\mathrm{CPP}i}\boldsymbol{\Delta}_{f_i}\boldsymbol{\Pi}^{l_i}\boldsymbol{\Lambda}_{c_1}^H\mathbf{F}^H\boldsymbol{\Lambda}_{c_2}^H\mathbf{x}+\widetilde{\mathbf{w}}\nonumber\\
	 	=&\sum_{i=1}^{P} h_i \mathbf{H}_i \mathbf{x}+ \widetilde{\mathbf{w}} \nonumber \\
	 	=&\mathbf{H}_{\mathrm{eff}}\mathbf{x}+\widetilde{\mathbf{w}}, 
	 	\end{align}
	 	where $\widetilde{\mathbf{w}}\sim \mathcal{CN}\left(\mathbf{0},N_0\mathbf{I}_N\right)$, $\boldsymbol{\Pi}$ is the  permutation (a.k.a. forward cyclic-shift) matrix, the Doppler shift matrix $\boldsymbol{\Delta}_{f_i}$ is a diagonal matrix given by   
	 	\begin{equation}
	 		\boldsymbol{\Delta}_{f_i} = \text{diag}\left(\left[1, e^{-j2\pi f_i}, \ldots, e^{-j2\pi f_i(N-1)^2}\right]^T\right),
	 	\end{equation}
	 	and the \emph{chirp} cyclic prefix $\boldsymbol{\Gamma}_{\mathrm{CPP}i}$ is an $N \times N$ diagonal matrix expressed as
	 	\begin{equation}
	 		\boldsymbol{\Gamma}_{\mathrm{CPP}i}= 
	 		\begin{cases}
	 			e^{-j2\pi c_1 \left( N^2-2N\left(l_i-n\right)\right)}, &n< l_i,\\
	 			1, & n\geq l_i.
	 		\end{cases}
	 	\end{equation}
	 	In (\ref{H_eff}), $\mathbf{H}_i(m,n)$ can be expressed as 
	 	\begin{equation}
	 		\mathbf{H}_i(m,n)=e^{j\frac{2\pi}{N}(Nc_1l_i^2-nl_i+Nc_2(n^2-m^2))}\delta\left(\left<n-m-\text{loc}_{i}\right>_N\right),
	 	\end{equation}
	 	where $\text{loc}_{i}\triangleq\alpha_{i}+2Nc_{1}l_{i}$, with  $\alpha_{i}=Nf_i\in [-\alpha_{\text{max}},\alpha_{\text{max}}]$ and $l_i \in [0,l_{\text{max}}]$ assumed to be integer valued numbers. Parameters $\alpha_{\text{max}}$ and $l_{\text{max}}$ are defined as the maximum Doppler shift index and maximum delay index, respectively. As proved in \cite{10087310}, 
	 	 in order to avoid different non-zero entries to coincide at the same position,
	 	$c_1$ should satisfy
	 	\begin{equation}
	 		c_1=\frac{2\alpha_{\text{max}}+1}{2N}.
	 	\end{equation}
	 	Parameter $c_2$ should be an arbitrary irrational number to achieve full diversity.
	\section{{Channel Estimation {and} Data Detection with Superimposed Pilots {for} AFDM}}
	\subsection{Superimposed Pilot Scheme}
	\begin{figure*}[tb]
		\center
		\includegraphics[width=7.0in]{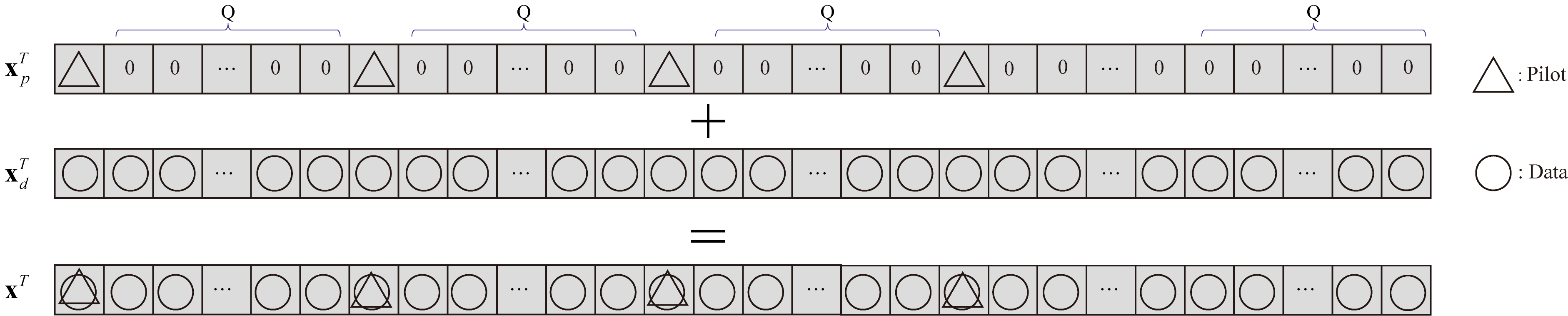}  
		\caption{Proposed superimposed pilot scheme. } \label{pilot}
$ $	\end{figure*}
	In the conventional embedded pilot scheme for AFDM, there are $Q = (l_{\text{max}}+1)(2\alpha_{\text{max}}+1)-1$ null guard samples placed on both sides of each pilot to ensure no interference between pilot and data signals, which results in a low SE. To increase the SE, we propose the  superimposed pilot scheme, whose structure is shown in Fig. \ref{pilot}. In our proposed scheme, we eliminate the guard samples to allow full data transmission and place the  pilots on top of the data signal in the DAFT domain. 
	  Therefore, the transmitting signal $\mathbf{x}$, formed by superimposing the pilot signal $\mathbf{x}_p=[x_{p0},x_{p1},\ldots,x_{p(N-1)}]^T \in \mathbb{C}^{N\times 1}$ and the data signal $\mathbf{x}_d=[x_{d0},x_{d1},\ldots,x_{d(N-1)}]^T \in \mathbb{C}^{N\times 1}$, is given by 
	\begin{equation}
		\mathbf{x}=\mathbf{x}_p +\mathbf{x}_d. 
	\end{equation}  
	 We define the pilot power as $\sigma_p^2$ and data power as $\sigma_d^2$, such that $\mathbf{x}_p^H\mathbf{x}_p = \sigma_p^2$ and $\mathbb{E}\{\mathbf{x}_d\mathbf{x}_d^H\} = \sigma_d^2\mathbf{I}_N$. Theoretically, we can insert full pilots over the whole DAFT domain, which means all elements of $\mathbf{x}_p$ are non-zero. However, as will be explained in the next subsection the interference between pilots will arise, which might deteriorate the channel estimation performance. Therefore, to circumvent this issue, we suggest to place multiple equally spaced pilots  with a spacing of at least $Q$, such that the element of $\mathbf{x}_p$ satisfies
	\begin{equation}
		x_{pm}=
		\begin{cases}
			\sigma_p/{\left(\sqrt{M+1}\right)}, &m\in\{0,Q+1,\ldots, M(Q+1)\}, \\
			0, &\text{otherwise},
		\end{cases}
	\end{equation}
	 where $ 0\leq M(Q+1)<N-Q$ and $M+1$ denotes the total number of superimposed pilots.  
	\subsection{Channel Estimation}
	Since there are at most $Q+1$ channel paths, the received signal can be written as 
	\begin{equation}\label{eq2}
		\mathbf{y}=\sum_{t=1}^{Q+1}h_t\boldsymbol{\Theta}_t(\mathbf{x}_p+\mathbf{x}_d)+\widetilde{\mathbf{w}},
	\end{equation}
	where $\boldsymbol{\Theta}_t=\boldsymbol{\Lambda}_{c_2}\mathbf{F}\boldsymbol{\Lambda}_{c_1}\boldsymbol{\Gamma}_{\mathrm{CPP}t}\boldsymbol{\Delta}_{f_t}\boldsymbol{\Pi}^{l_t}\boldsymbol{\Lambda}_{c_1}^H\mathbf{F}^H\boldsymbol{\Lambda}_{c_2}^H$. The values of the  delay index $l_t$ and the Doppler shift index $\alpha_t=Nf_t$  associated with index $t$ are given by
	\begin{equation}
		l_t= \left\lfloor\frac{t-1}{2Nc_1}\right\rfloor
	\end{equation}
	and
	\begin{equation}
		\alpha_t=\left<t-1\right>_{2Nc_1}-\alpha_{\text{max}},
	\end{equation}
	respectively. The received signal vector $\mathbf{y}$ in (\ref{eq2}) can be expressed in a more compact form as 
	\begin{equation}\label{eq3}
		\mathbf{y}=\boldsymbol{\Phi}_p\mathbf{h}+\boldsymbol{\Phi}_d\mathbf{h}+\widetilde{\mathbf{w}},
	\end{equation}
	where $\boldsymbol{\Phi}_p(\boldsymbol{\Phi}_d) \in \mathbb{C}^{N\times (Q+1)}$ and $\mathbf{h}\in\mathbb{C}^{(Q+1)\times 1}$. The channel state information (CSI) vector $\mathbf{h}$ has  a mean of  $\mathbb{E}\{\mathbf{h}\}=\mathbf{0}$ and a  covariance matrix of $\mathbf{C_h}=\mathbb{E}\{\mathbf{h}\mathbf{h}^H\}=\text{diag}([\sigma_{h1}^2,\sigma_{h2}^2,\ldots,\sigma_{h(Q+1)}^2]^T)$. The matrices $\boldsymbol{\Phi}_p$ and $\boldsymbol{\Phi}_d$, which contain the pilot vector $\mathbf{x}_p$ and the data vector $\mathbf{x}_d$, respectively, can be obtained as
	\begin{align}
		&\boldsymbol{\Phi}_p=\left[\boldsymbol{\Theta}_1 \mathbf{x}_p, \boldsymbol{\Theta}_2 \mathbf{x}_p, \ldots, \boldsymbol{\Theta}_{(Q+1)} \mathbf{x}_p \right],\\
		&\boldsymbol{\Phi}_d=\left[\boldsymbol{\Theta}_1 \mathbf{x}_d, \boldsymbol{\Theta}_2 \mathbf{x}_d, \ldots, \boldsymbol{\Theta}_{(Q+1)} \mathbf{x}_d\right]. 
	\end{align}
	
	{\it{Lemma}} 1: The mean and covariance matrix of the random matrix $\boldsymbol{\Phi}_d$ are $\mathbb{E}\{\boldsymbol{\Phi}_d\}=\mathbf{0}$ and $\mathbb{E}\{\boldsymbol{\Phi}_d\boldsymbol{\Phi}_d^H \}=\sigma_d^2(Q+1)\mathbf{I}_N$, respectively.
	 \begin{IEEEproof}
	 	See Appendix A.
	 \end{IEEEproof}
	The pilots and data are mixed in the received vector $\mathbf{y}$, causing the pilot-data interference. However, the data and pilots are statistically independent. Therefore, we can treat $\boldsymbol{\Phi}_d\mathbf{x}_d$ as part of the effective noise in the process of channel estimation. Furthermore, we define  $\widehat{\mathbf{w}}=\boldsymbol{\Phi}_d\mathbf{h}+\widetilde{\mathbf{w}}$ as the effective noise in the channel estimation.
	
	{\it{Lemma}} 2: The vector $\widehat{\mathbf{w}}$ has a mean of  $\mathbb{E}\{\widehat{\mathbf{w}}\}=0$ and a covariance matrix of 
	\begin{equation} \label{lemma2}
		\mathbf{C}_{\widehat{\mathbf{w}}}=\mathbb{E}\left\{\widehat{\mathbf{w}}\widehat{\mathbf{w}}^H\right\}=\left(\left(\sum_{i=1}^{Q+1}\sigma_{h_i}^2\right)\sigma_d^2+N_0\right)\mathbf{I}_N=\sigma_{\widehat{\mathbf{w}}}^2\mathbf{I}_N.
	\end{equation} 
	 
	\begin{IEEEproof}
		See Appendix B.
	\end{IEEEproof}
	
	Given the statistics of the channel and effective noise, the minimum mean square error (MSE) estimate $\widehat{\mathbf{h}}$ can be readily obtained as \cite{9539066}
		\begin{equation}
			\widehat{\mathbf{h}}=\left(\boldsymbol{\Phi}_p^H\mathbf{C}_{\widehat{\mathbf{w}}}^{-1}\boldsymbol{\Phi}_p+\mathbf{C_h}^{-1}\right)^{-1}\boldsymbol{\Phi}_p^H\mathbf{C}_{\widehat{\mathbf{w}}}^{-1}\mathbf{y}.
		\end{equation}
		
	   {\it{Lemma}} 3: To minimize the MSE  $\mathbb{E}\{||\widehat{\mathbf{h}}-\mathbf{h}||^2\}$, the column vectors of the matrix $\boldsymbol{\Phi}_p$ should form a set of orthogonal bases. This means for any integer $m,n \in [1,2,\ldots,Q+1]$, the following condition should hold 
	\begin{align}  
		\mathbf{x}_p^H\boldsymbol{\Theta}_m^H\boldsymbol{\Theta}_n\mathbf{x}_p= \label{basis1}
		\begin{cases}
			\sigma_p^2, &\text{if }  m=n, \\
			0, &\text{if } m\neq n.
		\end{cases}  		
	\end{align}
     This condition ensures that the columns of $\boldsymbol{\Phi}_p$ are orthogonal to each other. In our proposed superimposed pilot scheme, the interval of two adjacent pilots, namely $Q$, is carefully chosen such that all column vectors in $\boldsymbol{\Phi}_p$ are orthogonal to each~other.  
	
	\begin{IEEEproof}
		See Appendix C.
	\end{IEEEproof}
	To alleviate the interference for channel estimation resulting from the data and noise, we introduce a threshold criterion to estimate $\widehat{\mathbf{h}}$. Therefore, if we define a path indicator vector $\mathbf{b}=[b_1,b_2,\ldots,b_{(Q+1)}]^T$, where $b_t$ indicates whether a path associated with $\boldsymbol{\Theta_t}$ exists or not, we have
	\begin{equation}
		b_t=
		\begin{cases}
			1, &\text{if } \widehat{h}_t>\gamma, \\
			0, &\text{otherwise}.
		\end{cases}
	\end{equation}
	 The value of $\gamma$ influences the channel estimation performance. When $\gamma$ is relatively large, some existing channel paths with a marginal gain may be missed. On the contrary, when $\gamma$ is relatively small, some non-existent paths will be erroneously included due to data and noise. According to our extensive simulations, the optimal performance is achieved when $\gamma$ is set as $3\sqrt{{\sigma_{\widehat{\mathbf{w}}}^2}/\sigma_p^2}$.  This is in consistent with the result reported for OTFS channel estimation in \cite{8671740}.
	
	\subsection{Data Detection}
	After obtaining the estimated CSI vector $\widehat{\mathbf{h}}$, we can now cancel the interference caused by the pilot signal for data detection as
	\begin{equation}
		\mathbf{y}_d=\mathbf{y}-\widehat{\mathbf{H}}_\mathrm{{eff}}\mathbf{x}_p,
	\end{equation}
	where $\widehat{\mathbf{H}}_\mathrm{{eff}}=\sum_{t=1}^{Q+1}b_t\boldsymbol{\Theta}_t\widehat{\mathbf{h}}$. Assuming the channel estimation is perfect, the received signal associated with the data can be approximately derived as
	\begin{equation}
		\mathbf{y}_d=\widehat{\mathbf{H}}_\mathrm{{eff}}\mathbf{x}_d+\widetilde{\mathbf{w}}.
	\end{equation}
	Since the matrix $\widehat{\mathbf{H}}_{\mathrm{eff}}$ is typically sparse, we can adopt the message passing (MP) algorithm that offers low complexity. 
	\subsection{Iterative Channel Estimation and Data Detection}
	Due to the mutual interference between data and pilot signals, both the channel estimation and data detection are imprecise.  To achieve more precise estimation,  we resort to the iterative interference cancellation. Let $\mathbf{b}^i$, $\widehat{\mathbf{h}}^i$ and $\widehat{\mathbf{x}}_d^i$ denote the path indicator vector, channel estimation and data detection after the $i$-th iteration, respectively. In the $(i+1)$-th iteration, the CSI vector can be estimated by
	\begin{equation}
		\widehat{\mathbf{h}}^{i+1}=\left(\boldsymbol{\Phi}_p^H\mathbf{C}_{\widehat{\mathbf{w}}}^{-1}\boldsymbol{\Phi}_p+\mathbf{C_h}^{-1}\right)^{-1}\boldsymbol{\Phi}_p^H\mathbf{C}_{\widehat{\mathbf{w}}}^{-1}\left(\mathbf{y}-\widehat{\mathbf{H}}_\mathrm{{eff}}^i\mathbf{x}_d^i\right),
	\end{equation} 
	where $\widehat{\mathbf{H}}_\mathrm{{eff}}^i=\sum_{t=1}^{Q+1}b_t^i\boldsymbol{\Theta}_t\widehat{\mathbf{h}}^i$. We compare the elements of $\widehat{\mathbf{h}}^{i+1}$ with the threshold $\gamma$ to obtain  $\mathbf{b}^{i+1}$ and $\widehat{\mathbf{H}}_{\mathrm{eff}}^{i+1}$. Finally,  ${\mathbf{x}}_d$ can be estimated from $\mathbf{y}_d^{i+1}=\mathbf{y}-\widehat{\mathbf{H}}_\mathrm{{eff}}^{i+1}\mathbf{x}_p$ via the MP algorithm after the $(i+1)$-th iteration.
	\section{Simulation Results}
	  In this section, we consider the AFDM system with $N=512$ and quadrature PSK. The maximum Doppler shift index $\alpha_{\text{max}}=2$, and the maximum  delay index $l_{\text{max}}=2.$ The number of channel paths is $P=3.$ Each channel path has a different delay index and random Doppler index chosen from $[-\alpha_{\text{max}},\alpha_{\text{max}}].$ Each channel coefficient follows the distribution $\mathcal{CN}\left(0,{1}/{P}\right)$.  The noise variance is set to $N_0=1$ dBm.  The data signal-to-noise ratio (SNR) for transmission and pilot SNR for channel estimation are defined as  $\text{SNR}_d={\sigma_d^2}/{N_0}$ and $\text{SNR}_p={\sigma_p^2}/{N_0}$, respectively. Following \cite{10333832}, we set $\text{SNR}_p=50$ dB. The simulation results are obtained from more than $10^6$ independent channel realizations.
	  
	  \begin{figure}[tb]
	  	\center
	  	\includegraphics[width=3in]{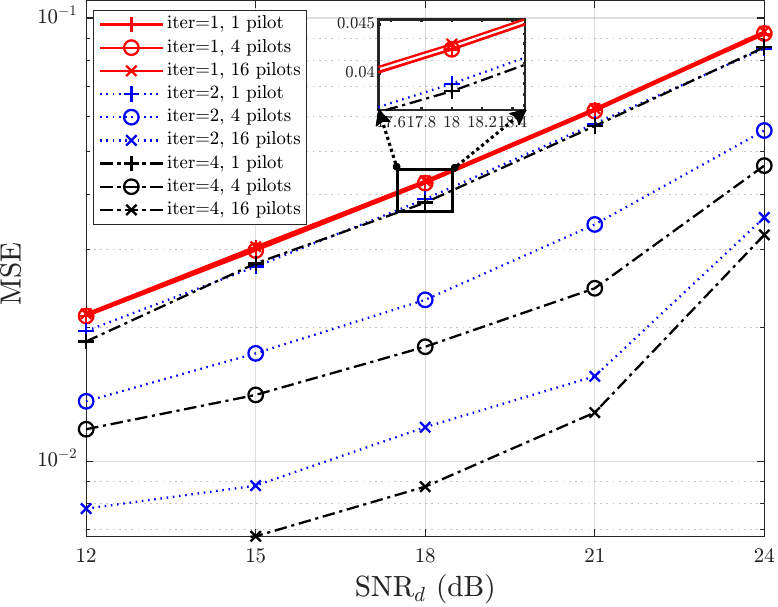}  
	  	\caption{MSE performance versus $\text{SNR}_d$ with $\text{SNR}_p=50$ dB. } \label{MSE}
	  \end{figure}
	  
	  In Fig. \ref{MSE}, we illustrate the MSE  performance. We consider the different numbers of iterations and superimposed pilots. It is observed that when the number of pilots is fixed, the MSE performance is improved as the number of iterations increases. This result confirms the effectiveness of our iterative algorithm. Furthermore, the increasing MSE performance with 4 and 16 pilots are more noticeable than that with a single pilot. In other words, as the pilot number increases, the channel estimation performs better and the effect of the iterative algorithm is more obvious. The results confirm multiple superimposed pilots with a reasonable placement achieve better performance in channel estimation. 
	  \begin{figure}[tb]
	  	\center
	  	\includegraphics[width=3in]{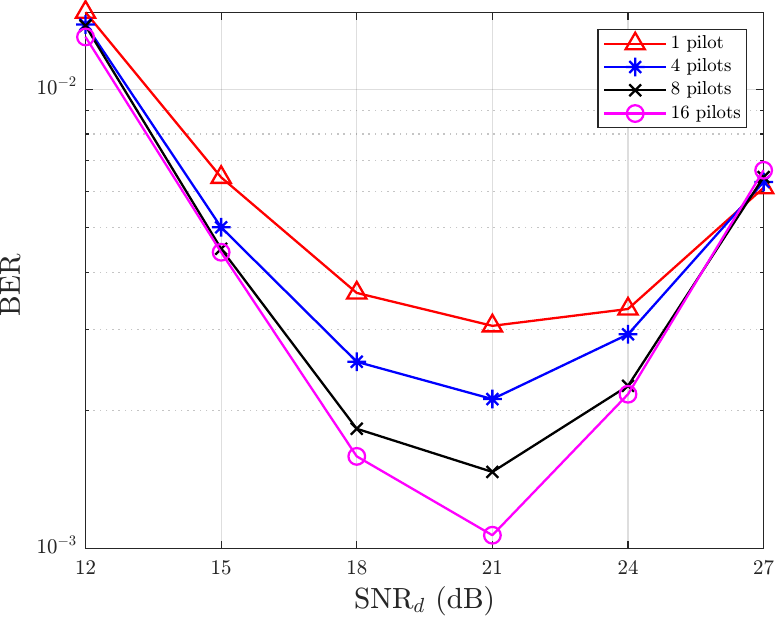}  
	  	\caption{BER performance versus $\text{SNR}_d$ with 2 iterations and $\text{SNR}_p=50$~dB.} \label{BER}
	  \end{figure}
	  
	  In Fig. \ref{BER}, we present the uncoded bit error rate (BER) performance based on the MP detector with 2 iterations.  Two interesting findings emerge from the figure. Firstly, as the number of pilots increases, the  minimum BER for each curve decreases when $\text{SNR}_d=$ 21 dB. This further confirms the benefit of our superimposed pilot scheme, and the fact there is an optimal power allocation for the pilot and data signals. Secondly, for each BER curve, as the $\text{SNR}_d$ increases for the range of $\text{SNR}_d$~$<21 $ dB, the BER decreases because the noise effect decreases. However, as the $\text{SNR}_d$ increases when $\text{SNR}_d$~$>21$~dB, the BER performance deteriorates with the increasing $\text{SNR}_d$. This phenomenon occurs because the pilot signal power is fixed. As the $\text{SNR}_d$ increases, there is more interference between the data and pilot signals which causes the inferior channel estimation presented in Fig.~\ref*{MSE}, leading to the degraded BER performance in Fig. \ref{BER}. 
	\section{Conclusion}
	   We have proposed an AFDM channel estimation scheme with superimposed pilots, aiming to increase the SE by superimposing data and pilots in the DAFT domain as well as to  minimize the channel estimation MSE by ensuring at least $Q$ guard intervals between two adjacent pilots. We have shown through rigorous derivations that our proposed scheme can effectively minimize the channel estimation MSE when the pilot signal power is fixed. Simulation  results  have demonstrated that both the MSE and BER performance  are improved as the number of pilots increases. For future work,
	   we will extend our proposed scheme to multi-input-multi-output and/or multi-user systems.

	\section*{APPENDIX A}
	 The random matrix $\boldsymbol{\Phi}_d$ satisfies	  
	\begin{equation}
		\mathbb{E}\left\{\boldsymbol{\Phi}_d\right\}=[\boldsymbol{\Theta}_1 \mathbb{E}\left\{\mathbf{x}_d\right\}\,\boldsymbol{\Theta}_2\mathbb{E}\left\{\mathbf{x}_d\right\}\,\cdots \,\boldsymbol{\Theta}_{Q+1}\mathbb{E}\left\{\mathbf{x}_d\right\} ]=\mathbf{0}
	\end{equation}
	and 
	\begin{equation}
		\mathbb{E}\left\{\boldsymbol{\Phi}_d\boldsymbol{\Phi}_d^H\right\}=\sum_{t=1}^{Q+1}\boldsymbol{\Theta}_t\mathbb{E}\left\{\mathbf{x}_d\mathbf{x}_d^H\right\}\boldsymbol{\Theta}_t^H=(Q+1)\sigma_d^2\mathbf{I}_N
	\end{equation}
	where $\boldsymbol{\Theta}_t\boldsymbol{\Theta}_t^H=\mathbf{I}_N.$
	\section*{APPENDIX B}
	Since the data signal is statistically independent of the noise $\widehat{\mathbf{w}}$, we can express the covariance matrix of $\widehat{\mathbf{w}}$ as 
	\begin{equation} \label{appendix1}
		\mathbb{E}\left\{\widehat{\mathbf{w}}\widehat{\mathbf{w}}^H\right\}=\mathbb{E}\left\{\boldsymbol{\Phi}_d\mathbf{h}\mathbf{h}^H \boldsymbol{\Phi}_d^H\right\}+\mathbb{E}\left\{\widetilde{\mathbf{w}}\widetilde{\mathbf{w}}^H\right\}.
	\end{equation} 
	 As proved in\cite{8752008}, if a random matrix $\mathbf{A}\in \mathbb{C}^{m\times n}$ has the covariance $\mathbb{E}\{\mathbf{AA}^H\}=\sigma \mathbf{I}_M$, for any Hermitian matrix $\mathbf{B} \in \mathbb{C}^{n\times n}$ it follows that $\mathbb{E}\{\mathbf{ABA}^H\}=\frac{\text{Tr}\left\{\mathbf{B}\right\}}{n}\mathbb{E}\{\mathbf{AA}^H\}$.
	Since $\mathbb{E}\left\{\boldsymbol{\Phi}_d\boldsymbol{\Phi}_d^H\right\}=(Q+1)\sigma_d^2\mathbf{I}_N$, the first term of the right hand side of (\ref{appendix1}) can be calculated as
	\begin{align} \label{appendix2}
		\mathbb{E}\left\{\boldsymbol{\Phi}_d\mathbf{hh}^H\boldsymbol{\Phi}_d^H\right\}&=\mathbb{E}\left\{\mathbb{E}\left\{\boldsymbol{\Phi}_d\mathbf{hh}^H\boldsymbol{\Phi}_d^H\right\}|\mathbf{h}\right\}\nonumber\\
		&=\sigma_d^2 \mathbb{E}\left\{\text{Tr}\left\{\mathbf{hh}^H\right\}|\mathbf{h}\right\}\mathbf{I}_N \nonumber \\
		&= \left(\sum_{i=1}^{Q+1}\sigma_{h_i}^2\right)\sigma_d^2\mathbf{I}_N.		
	\end{align}
	Substituting (\ref{appendix2}) and the covariance matrix of $\widehat{\mathbf{w}}$ into (\ref{appendix1}), we arrive at (\ref{lemma2}).
	\section*{APPENDIX C}
    The covariance matrix of $\widehat{\mathbf{h}}$ can be calculated as \cite{kay1993fundamentals} 
    \begin{equation} \label{cov}
    	\mathbf{C}_{\widehat{\mathbf{h}}}=\left(\boldsymbol{\Phi}_p^H\mathbf{C}_{\widehat{\mathbf{w}}}^{-1}\boldsymbol{\Phi}_p+\mathbf{C_h}^{-1}\right)^{-1},
    \end{equation}
    where the second term  can be omitted as $\sigma_p^2 \gg 1$. The MSE of the channel estimation is thereby given by
    \begin{align}
    	\mathbb{E}\left\{\left\|\widehat{\mathbf{h}}-\mathbf{h}\right\|^2\right\}&=\text{Tr}\left\{\mathbf{C}_{\widehat{\mathbf{h}}}\right\}\nonumber\\
    	&\approx \text{Tr} \left\{\left(\boldsymbol{\Phi}_p^H\mathbf{C}_{\widehat{\mathbf{w}}}^{-1}\boldsymbol{\Phi}_p\right)^{-1}\right\}\nonumber\\
    	&=\sigma_{\widehat{\mathbf{w}}}^2 \text{Tr}\left\{\left(\boldsymbol{\Phi}_p^H\boldsymbol{\Phi}_p\right)^{-1}\right\}.
    \end{align}
     Note that the trace of $\boldsymbol{\Phi}_p^H\boldsymbol{\Phi}_p$ is given by
    \begin{equation} \label{trace}
    	\text{Tr}\left\{\boldsymbol{\Phi}_p^H\boldsymbol{\Phi}_p\right\}=\sum_{t=1}^{Q+1}\mathbf{x}_p^H\boldsymbol{\Theta}_t^H\boldsymbol{\Theta}_t\mathbf{x}_p=(Q+1)\sigma_p^2.
    \end{equation}
      Also note that for a square matrix $\mathbf{A}$, the trace of $\mathbf{A}$ is equal to the sum of its eigenvalues, namely $\text{Tr}\{\mathbf{A}\}=\sum_{i}\varepsilon_i$, where $\varepsilon_i$ is the $i$-th eigenvalue of $\mathbf{A}$. If $\mathbf{A}$ is invertible, then $\{1/\varepsilon_i\}$ are the eigenvalues of $\mathbf{A}^{-1}$. We define~$\lambda_l, l\in \{1,2,\ldots,Q+1\}$ as the eigenvalue of $\boldsymbol{\Phi}_p^H\boldsymbol{\Phi}_p$.   According to (\ref{trace}), the sum of the eigenvalues $\sum_l \lambda_l$ is fixed.  Minimizing the trace of $(\boldsymbol{\Phi}_p^H\boldsymbol{\Phi}_p)^{-1}$ is equivalent to minimizing $\sum_l 1/\lambda_l$. Therefore, all the eigenvalues $\{\lambda_l\}$ need to be identical. Since $\boldsymbol{\Phi}_p^H\boldsymbol{\Phi}_p$ has equal diagonal elements, all the column vectors in $\boldsymbol{\Phi}_p^H\boldsymbol{\Phi}_p$ should be orthogonal to each other.
     Let us consider two arbitrary column vectors $\boldsymbol{\Theta}_j\mathbf{x}_p$ and $\boldsymbol{\Theta}_i\mathbf{x}_p$ in $\boldsymbol{\Phi}_p$, where $i,j\in [1,Q+1] $. The ($m$,$n$)-th entry of $\boldsymbol{\Theta}_i$ can be expressed~as
     \begin{equation}
     	\boldsymbol{\Theta}_i\left(m,n\right)=e^{j\frac{2\pi}{N}(Nc_1 l_i^2-n l_i +Nc_2(n^2-m^2))} \delta(\left<n-m-\text{loc}_i\right>_N).
     \end{equation} 
     Therefore, $\boldsymbol{\Theta}_j^H\boldsymbol{\Theta}_i\left(m,n\right)$ can be written as 
     \begin{align}
     	\boldsymbol{\Theta}_j^H\boldsymbol{\Theta}_i\left(m,n\right)&=\sum_{k=1}^{N} \boldsymbol{\Theta}_j^H\left(m,k\right) \boldsymbol{\Theta}_i\left(k,n\right) \nonumber \\
     	&=\zeta_{m,n} \delta\left(\left<n-m-\text{loc}_i+\text{loc}_j\right>_N\right)
     \end{align}
     where
     \begin{equation}
     	\zeta_{m,n}=e^{j\frac{2\pi}{N}(Nc_1 \left(l_i^2-l_j^2\right)-\left(n l_i-ml_j \right)+Nc_2(n^2-m^2))}.
     \end{equation}  
     On the other hand, the inner product between $\boldsymbol{\Theta}_j\mathbf{x}_p$ and $\boldsymbol{\Theta}_i\mathbf{x}_p$ can be expressed as 
     \begin{equation} \label{appendix3}
     	\mathbf{x}_p^H\boldsymbol{\Theta}_j^H\boldsymbol{\Theta}_i\mathbf{x}_p=\sum_{m=0}^{N-1}x_{pm}^*x_{p\left<m+\text{loc}_i-\text{loc}_j\right>_N}\zeta_{m,\left<m+\text{loc}_i-\text{loc}_j\right>_N}.
     \end{equation}
     Since  $\text{loc}_i-\text{loc}_j \in [-Q,Q]$ and  $Q$ null guard samples are placed on both sides of each non-zero $x_{pm}$ in $\mathbf{x}_p$,  the value of $x_{pm}^*x_{p\left<m+\text{loc}_i-\text{loc}_j\right>_N}$ is always zero when $i\neq j$. Equation~(\ref{appendix3}) can be thereby simplified as $\mathbf{x}_p^H\mathbf{x}=\sigma_p^2$  when $i=j$.

\bibliographystyle{IEEEtran}
\bibliography{reference}
\end{document}